\documentclass[reprint, superscriptaddress, prl]{revtex4-1}

\usepackage{amsmath, amssymb, mathrsfs}
\usepackage{mathtools}
\usepackage{slashed}
\usepackage[english]{babel}
\usepackage{graphicx,color,psfrag}
\usepackage{placeins}

\newcommand{\lag}{\mathcal{L}_{eff}}
\newcommand{\borelag}{\mathcal{BL}_{eff}}

\renewcommand{\Im}{\mathfrak{Im}}
\newcommand{\dd}{\mathrm{d}}

\begin{document}
\author{Adrien Florio}
\affiliation{Institute of Physics, Laboratory of Particle Physics and Cosmology (LPPC), \'Ecole Polytechnique F\'ed\'erale de Lausanne (EPFL), CH-1015 Lausanne, Switzerland.}
\date{December 27, 2019}
\title{Schwinger pair production from Pad\'e-Borel reconstruction}

\begin{abstract}
In this work, we show how the knowledge of the first few terms of the Euler-Heisenberg Lagrangian's weak-field expansion in a magnetic field background is enough to reconstruct the pair-production rate in a strong electric field background. To this end, we study its associated truncated Borel sum using Pad\'e approximants, as advocated in a recent work by Costin and Dunne, J. Phys. A52, 445205 (2019).
\end{abstract}
\maketitle

\section{Introduction}
In recent years, the program of "resurgence" has started to collect a number of successes in quantum mechanics and field theory.
The idea behind it is that the typical asymptotic expansions that are to be dealt with, for example usual weak coupling expansions, are to be understood as being part of a transseries. In simple terms, transseries are sums of asymptotic series weighted by non-perturbative factors such as exponentials and logarithms. A typical example is the semi-classical expansion, which is a sum of perturbative/asymptotic expansions around different saddle points. We refer the reader to \cite{costin2008asymptotics,Marino:2012zq} for pedagogical introductions to the topic.

The very analytic structure of transseries implies consistency relations between the different constituent asymptotic series. In particular, large order coefficients of a given expansion are known to be related to the small order coefficients of neighboring expansions. While being seemingly a mathematical curiosity, these relations have, for example, been used to predict the loop expansion around an instanton background for the quantum mechanical  Sine-Gordon potential \cite{Dunne:2014bca}, predictions which have been explicitly verified up to three loops using diagrammatic methods \cite{Escobar-Ruiz:2015nsa}. For other interesting examples and reviews, we defer the reader to \cite{Dunne:2016jsr,Ahmed:2017lhl,Aniceto:2018bis,Marino:2019wra,Gukov:2016njj} and references therein.

An immediate complaint against the potential practical usefulness of such approaches is that the knowledge of large orders terms of realistic quantum field theories expansions is not necessarily available. In this spirit, reference~\cite{Costin:2019xql} started to investigate the amount of "non-perturbative" information that can be extracted from a finite number of terms of an asymptotic expansion.  Stunningly, using relatively few terms of the asymptotic expansion of solutions to the Painlev\'e I equation around real infinity, they were able to reconstruct the whole highly non-trivial analytic structure of this solution throughout the whole complex plane. In a similar spirit, the works \cite{Serone:2018gjo,Serone:2019szm} successfully explored the phase diagram of the $\lambda\phi^4$ field theory by computing weak coupling expansions up to nine loops and studying their associated Borel sums.

Having in mind general quantum field theories, these are proofs of principle that a lot of non-perturbative information might be at our hand, waiting to be extracted from perturbative expansions.

This short note's aim is to illustrate again the potential use of some of the ideas developed in these works in field theory, using one of the simplest "non-perturbative" effects at hand, namely Schwinger pair production. In particular, we will present two results. First, the knowledge of a few terms of the weak-field expansion of the Euler-Heisenberg effective Lagrangian in a background magnetic field is enough to reconstruct its strong-field behavior. Then, and perhaps more interestingly, the same knowledge is enough to reconstruct the Euler-Heisenberg effective Lagrangian in a background electric field, for weak and strong fields, including its imaginary part. This means that this imaginary part, which gives the particle production rate in a constant electric field, can be inferred from a few terms of a perturbative expansion.


\section{Schwinger effect, generalities}
Schwinger pair production is one of the most basics field-theoretic non-perturbative effect, see \cite{Dunne:2004nc} for an extensive review. Its simplest realization is the vacuum emission of charged particles in the presence of strong electric fields. A way to study it is to compute the one-loop fermionic effective action in a background electromagnetic field. Then, the phenomenon of pair-productions is signaled by the appearance of an imaginary part in the effective action. For the sake of simplicity, we will hereafter restrict ourselves to the constant background case. There, one can explicitly write down the effective Lagrangian \cite{Heisenberg:1935qt}. For a purely magnetic field, it admits the following closed-form \cite{Dunne:2004nc}
\begin{align}
	&\lag(B) = \frac{(eB)^2}{2 \pi^2}\bigg [  \zeta'_H\left(-1,\frac{m^2}{2 e B}\right) \label{eq:closedLB}\\
	&+ \zeta_H\left ( -1,\frac{m^2}{2 e B}\right)\ln\left(\frac{m^2}{2 e B}\right)-\frac{1}{12}+\frac{1}{4}\left(\frac{m^2}{2 e B}\right)^2\bigg ] ,\notag
\end{align}
with $\zeta_H(s,a)$ the Hurwitz zeta function and $\zeta'_H(s,a)$ its derivative with respect to $s$. The parameters $m$ and $e$ are respectively the fermion's mass and electric charge, while $B$ is the strength of the constant background magnetic field. This expression is real and there is no pair-production in a magnetic background, as there is apriori no magnetically charged particle to be produced. The case of a pure electric field background is recovered by analytically continuing $B\to \pm i E$ \cite{Costin:2017ziv} (note that in this sense \eqref{eq:closedLB} can also be understood as the Euclidean space effective Lagrangian in an electric field background). Then, the effective Lagrangian does develop an imaginary part, which can be written as \cite{Costin:2017ziv} 
\begin{align}
\Im\left(\lag(E)\right) &= \frac{m^4}{8 \pi^3}\left(\frac{e E}{m^2}\right)^2\mathrm{Li_2}\left ( e^{-\frac{m^2\pi}{eE}} \right)\\
&= \frac{m^4}{8 \pi^3}\left(\frac{e E}{m^2}\right)^2 \left [e^{-\frac{m^2\pi}{eE}}+ \dots\right ],
\label{eq:imPart}
\end{align}
with $\mathrm{Li_2}$ the second polylogarithm. From this expression, it is easy to see the famous exponential suppression to the production rate $\Gamma_{prod}$, which by definition is \cite{Dunne:2004nc}
\begin{equation}
	\Gamma_{prod} = 2 \Im\left(\lag\right)\label{eq:rateDef}.
\end{equation}
Another representation of \eqref{eq:closedLB} that will be of use is the following Laplace-type integral \cite{Heisenberg:1935qt,Schwinger:1951nm}

\begin{align}
	\lag(B)=&-\frac{e^2B^2}{8 \pi^2}	\label{eq:anBorelSum}\\
	&\int_0^\infty \frac{\dd p}{p^2}\left(\coth{p}-\frac{1}{p}-\frac{p}{3}\right)e^{-\frac{m^2 p}{eB}}\notag.
\end{align}
From this representation, it is clear that the imaginary part in the electric case  comes from the contribution to the integral of the poles of the hyperbolic cotangent at integer multiples of $i\pi$.

In the rest of this work, we will be concerned with the weak-field expansion of \eqref{eq:closedLB}. It is given as \cite{Dunne:2004nc,Chadha:1977my}

\begin{align}
	\lag(B)&\sim\frac{m^4}{4\pi^2}\label{eq:weakB}\\
	&\sum_{n=0}^\infty(-1)^{n}(2n+1)!\frac{\zeta(2n+4)}{(2\pi)^{2n+4}}\left(\frac{2 e B}{m^2}\right)^{2n+4},\notag
\end{align}
with $\zeta(x)$ the Riemann zeta function. For the electric field, the expansion reads

\begin{align}
	\lag(E)&\sim\frac{m^4}{4\pi^2}\label{eq:weakE}\\
	&\sum_{n=0}^\infty(2n+1)!\frac{\zeta(2n+4)}{(2\pi)^{2n+4}}\left(\frac{2 e E}{m^2}\right)^{2n+4}.\notag
\end{align}

Both series are asymptotic because of the factorial growth of their coefficients. They are also both real to all orders. It is in this sense that the rate \eqref{eq:rateDef} is a non-perturbative quantity; at any given order in \eqref{eq:weakE}, $\Gamma_{prod}=0$.

Actually, these weak-field expansions can be resummed to \eqref{eq:anBorelSum} using Borel summation. In this language, again, the imaginary part appears because of the presence of poles in the Laplace transform; \eqref{eq:weakB} is "Borel summable" while \eqref{eq:weakE} is not.

%
%
%
%
%

\section{Strong-Field Regime From Weak-Field Expansion}

Following \cite{Costin:2019xql}, we want to understand how much of the full Lagrangian \eqref{eq:closedLB} we can reconstruct using a finite number of terms in \eqref{eq:weakB}. To this purpose, again as in \cite{Costin:2019xql},  we construct the corresponding truncated Borel sum. From it, we build Pad\'e approximants, which are then used to compute a resummed Lagrangian $\lag^{res}$ through a Laplace transform. The idea behind this procedure is to try to exploit the fact that, while the original expansion is only asymptotic, its Borel transform is convergent. Note also that very similar techniques were already used in the $80's$, see for example \cite{Drouffe:1983fv} for a thorough review on QCD strong coupling expansion.

To keep notations clear, we set $x=\frac{m^2}{2 e B}$ and write the asymptotic expansion \eqref{eq:weakB}, truncated at order $N$, as

\begin{align}
\frac{\lag(x,N)}{m^4}&\sim\frac{1}{64\pi^6}\frac{1}{x^4}\\
&\sum_{n=0}^N(-1)^{n}(2n+1)!\frac{\zeta(2n+4)}{(2\pi)^{2n}}\left(\frac{1}{x}\right)^{2n}\notag\\
&=\frac{1}{64\pi^6}\frac{1}{x^4}\sum_{n=0}^N a_{2n}\left(\frac{1}{x}\right)^{2n}\label{eq:asX},
\end{align}
with $a_{2n}=(-1)^{n}(2n+1)!\frac{\zeta(2n+4)}{(2\pi)^{2n}}$. We also define a truncated Borel sum
\begin{align}
	\borelag(p,N)=&\sum_{n=1}^N \frac{a_{2n}}{(2n-1)!}p^{2n-1}.\label{eq:truncborel}
\end{align}

With these definitions, we construct a Pad\'e approximant of \eqref{eq:truncborel}. Pad\'e approximants are rational functions constructed to match a given series at specific points. They are typically used to try to reproduce the analytical structure of a function by extrapolating it away from some regions. Their rational nature allows for the emergence of poles and  branch cuts, which appear as accumulations of poles. They can be found in a variety of places in the physics literature. As a specific example, we can mention attempts to analytically continue Euclidean lattice data to Minkowski space through Pad\'e approximants, see \cite{Tripolt:2018xeo} and references therein.

	To have easy access to the poles of our Pad\'e function and have good control over the numerical Laplace transform, we use Pad\'e approximants of the type
\begin{align}
	\mathcal{P}^{2N} \borelag(p,N) = \sum_{n=1}^{N} \frac{c_n}{1+ b_n p}.\label{eq:padeapprox}
\end{align}
The coefficients $c_n, b_n$, which are in principle complex numbers, are computed by matching this expression to \eqref{eq:truncborel} around $p=0$, see \cite{BREZINSKI201569} for an explicit algorithm.

Finally, we compute our resummed Lagrangian as follows
\begin{align}
	\frac{\lag^{res}(x,N)}{m^4} &= \frac{1}{64\pi^6}\frac{1}{x^4}\label{eq:truncRes}\\
	&\left(a_0 + \int_0^\infty \dd p e^{-p x} \mathcal{P}^{2N} \borelag(p,N) \right).\notag
\end{align}
Note in particular that without the Pad\'e interpolation, we would have achieved nothing, as in this case  \eqref{eq:truncRes} would literally be equal to \eqref{eq:asX}.

\begin{figure}
\centering
\includegraphics{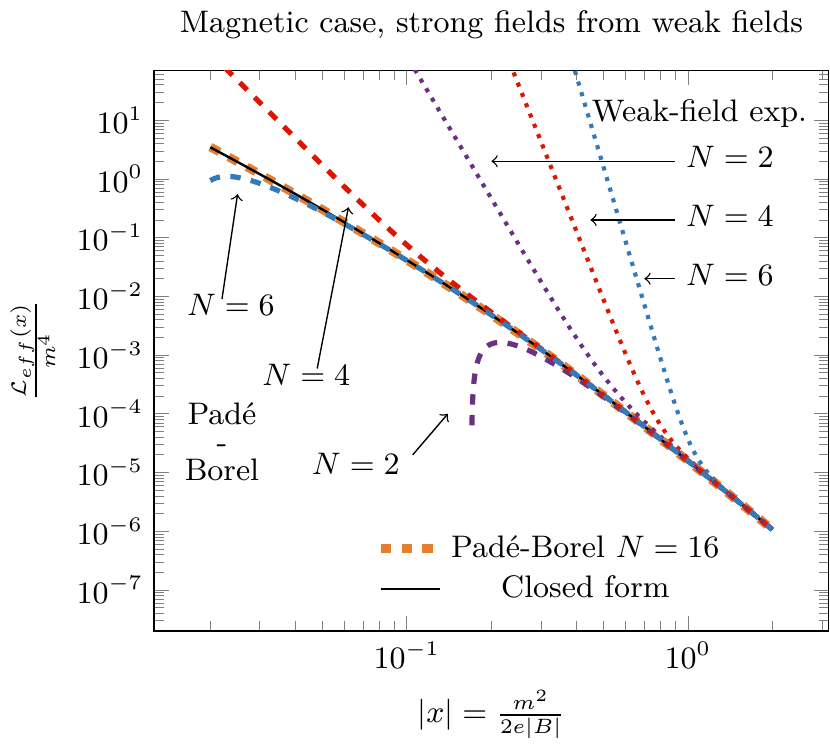}
\caption{
	Magnetic field effective Lagrangian. Closed-form (plain line), weak-field expansion (dotted lines) and Pad\'e-Borel reconstruction (dashed lines) for different truncation order $N$. The weak-field expansion has a typical asymptotic behavior; every order makes it break down faster. The Pad\'e-Borel reconstruction takes advantage of the fact that the Borel sum is convergent; every order improves the answer.
 }
 \label{fig:strongfromweak}
\end{figure}

We show  the result of this procedure, which from now on we will refer to as Pad\'e-Borel reconstruction, in figure \ref{fig:strongfromweak}. The plain black line is the closed-form \eqref{eq:closedLB}. The dotted lines are the truncated weak-field expansions, for different truncation $N$. The dashed lines are the Pad\'e-Borel reconstructed expressions for the same $N$. Note that $x\to \infty$ resp. $x\to 0$ corresponds to the weak resp. strong-field regime, the goal being to be able to extrapolate from the former to the latter. Being an asymptotic expansion, every order makes it break down for larger values of $x$, i.e. for weaker fields. On the contrary, the Pad\'e-Borel reconstruction improves as $N$ increases. This boils down to the fact that the Borel sum \eqref{eq:truncborel} is convergent; every new order contributes making the result more accurate. For example, only four terms of the weak-field expansion can be used to probe the strongly-coupled regime as far as $x=0.2$.

This is our first result. With the knowledge of only the first few terms of the weak-field expansion \eqref{eq:weakB}, it is possible to explore the regime of strong magnetic fields by first constructing the corresponding truncated Borel sum, Pad\'e approximating it and computing its Laplace transform.

\section{Schwinger effect reconstructed}

Now, we will show that this method, using the same data, actually also gives access to the regime of strong electric fields. In particular, we will see that we can use it to recover the Schwinger pair production rate.

To consider an electric field, we proceed with the analytic continuation $x\to \mp i x$. This leads us to study

\begin{align}
	\frac{\lag^{res}(\mp i x,N)}{m^4} &= \frac{1}{64\pi^6}\frac{1}{x^4}\\
	&\left(a_0 + \int_0^\infty \dd p e^{\pm i p x} \mathcal{P}^{2N} \borelag(p,N) \right).\notag
\end{align}

Technically, to compute this Laplace transform, we consider all the different fractions of \eqref{eq:padeapprox} separately. We then rotate  the integration contour in the complex plane by some angle and take into account any poles we might have crossed in the process.

\begin{figure}
\includegraphics{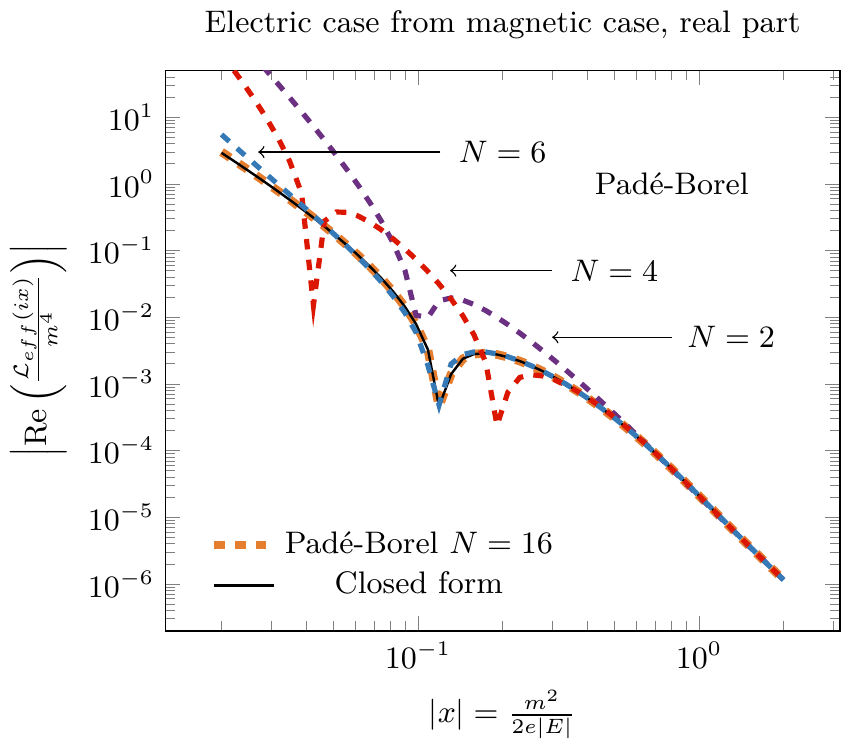}
\caption{	Real part of electric field effective Lagrangian. Closed-form (plain line) and Pad\'e-Borel reconstruction (dashed lines) for different truncation order $N$. The Pad\'e-Borel reconstruction leads to correct and convergent results even after analytic continuation.}
\label{fig:realE}
\end{figure}

Let us first look at what we obtain for the real part of the resummed electric field  effective Lagrangian obtained through this analytic continuation, figure \ref{fig:realE}. As in the magnetic case, few terms of the weak Lagrangian allows for a precise extrapolation up into the strong-field regime. In particular, the reconstruction is able to predict correctly non-trivial features such as the change of signs which happens around $x=0.1$ (note that we are plotting the absolute value).

\begin{figure}
\includegraphics{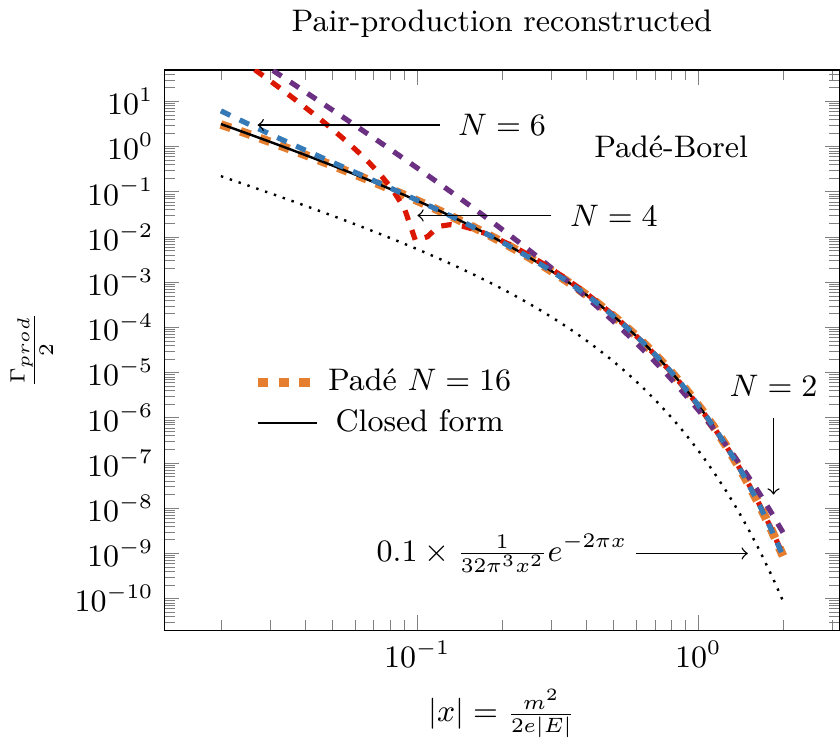}
\caption{Pair-production rate in a background electric field (imaginary part of the electric field effective Lagrangian). Closed-form (plain line) and Pad\'e-Borel reconstruction (dashed lines) for different truncation order $N$. The dotted line is the leading exponential suppression to the Schwinger rate (shifted for readability). While $N=2$ gives an imaginary part which is only qualitatively correct for weak-field, $N=4$ and larger leads to a quantitatively correct prediction of the Schwinger rate for a whole range of field-strengths.}
\label{fig:imagE}
\end{figure}

More interesting are the results for the imaginary part of the effective Lagrangian, i.e. the pair-production rate. They are shown in figure \ref{fig:imagE}. They behave in exactly the same way; few terms of the weak-field expansion still give a quantitatively correct prediction of the rate. As little as the first two terms are required to reconstruct an imaginary part which is qualitatively correct at weak-field. With only the first six terms one can make quantitative predictions up to strong fields. This has to be contrasted again with the original asymptotic series, which uses the same data but is real to all orders.

The perhaps surprising capability of the Pad\'e-Borel reconstruction to recover the pair-production rate is due to the fact that the Pad\'e approximants of the truncated Borel sums are able to reproduce the correct analytic structure of the Borel sum. In terms of our variable $x$, the actual Borel sum \eqref{eq:anBorelSum} is a meromorphic function with single poles at $x=2\pi i n$ for $n\in\mathbf{Z}, n\neq 0$. As already mentioned, the imaginary part \eqref{eq:imPart} can be understood as coming from the contribution of every single pole. It is dominated by the lowest-lying ones at $ x=\pm2\pi i$
\begin{align}
	\frac{\Gamma_{prod}^{lead.}}{2}=\frac{1}{32\pi^3 x^2}e^{-2\pi x},
\end{align}
which we also show in figure \ref{fig:imagE}.

\begin{figure}
	\includegraphics{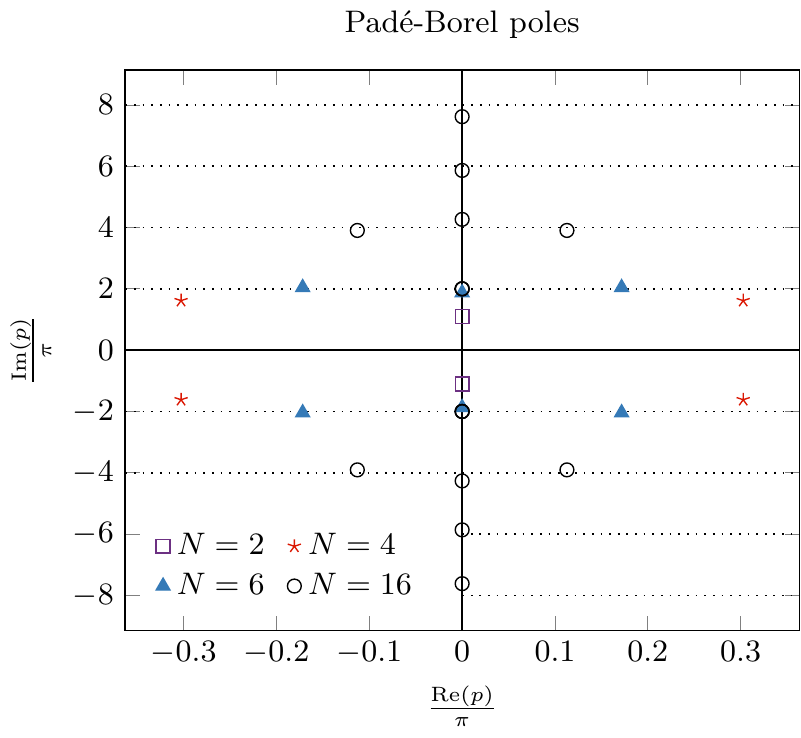}
	\caption{Poles of the Pad\'e-Borel reconstruction in the Borel plane, for different truncation order $N$. Dotted lines are multiples of $2\pi i$, where poles accumulate as $N$ is taken larger. This is the correct analytic structure of the actual Borel sum, which has single poles at non-zero multiples of $2\pi i$. Note that the Pad\'e-Borel approximation requires more than a single pole per multiple of $2\pi i$ to reproduce the correct functional dependence.}
	\label{fig:padePoles}
\end{figure}

As the Pad\'e-Borel approximants are constructed only from an asymptotic expansion around the real axis it is, however, a non-trivial fact that they are able to mimic correctly this analytic structure.
We show it occurring in figure \ref{fig:padePoles}, where we display the poles of our Pad\'e approximants. As the truncation order $N$ is taken to be larger, they accumulate around $x=2\pi i n$. Note that to approximate the correct prefactors, a single pole is replaced by a combination of different ones centered around $x=2\pi i n$. The leading poles at $\pm 2\pi i$ are first reproduced accurately by the truncation order $N=6$, which is consistent with the behavior of the results presented in figure \ref{fig:imagE}.

This is our second and most important result. The knowledge of a few terms of the weak-field expansion of the effective Euler-Heisenberg Lagrangian in a magnetic field background is enough to reconstruct the particle production rate in a strong electric field.

\section{Conclusion}
This work can be summarized as follows: using only the truncated weak-field asymptotic expansion of the Euler-Heisenberg effective Lagrangian in a magnetic background, we were able to reconstruct the full Euler-Heisenberg Lagrangian, including its imaginary part which gives the Schwinger pair production rate. This may come as a surprise, as this rate is zero at all-orders of the weak-field asymptotic expansion.

What this result suggests, as already realized in \cite{Costin:2019xql}, is that all coefficients in such asymptotic expansions contain information about the analytic structure of the underlying transseries in the whole complex plane.  This information can be extracted by studying the associated Borel sum even upon truncation to a finite number of terms, by taking advantage of the fact that the Borel sum is a  convergent series. Another remarkable fact is that the knowledge of the truncated Borel sum along the real axis is enough to gain information about its analytic structure, using  Pad\'e approximants, throughout the complex plane. In particular, this means that the underlying transseries is constrained enough to force the Pad\'e approximants to develop poles at the correct locations.

The precise mechanism behind this phenomenon remains to be better understood; this will be essential to apply this method to unsolved problems and obtain trustworthy predictions. The  aim of this work was however to illustrate its potential use in a simple physical problem.

\textit{Acknowledgments.} The author wants to thank G.~Dunne for an inspiring talk, S.~Valgushev for stimulating conversations and its invitation to BNL, where the aforementioned talk was given, and M.~Shaposhnikov for feedback on this work. The author is supported by the Swiss National Science Foundation.

%


\end{document}